\documentstyle[12pt]{article}
\begin{document}

\title{A Classification and Analysis of Higgs-flavor Models}
\author{{\bf S.M. Barr} \\
{\small Department of Physics and Astronomy} \\ {\small Bartol
Research Institute,}
\\ {\small University of Delaware, Newark, Delaware 19716} \\ \\
{\bf Thomas W. Kephart} \\ {\small Department of Physics and Astronomy,} \\
{\small Vanderbilt University, Nashville, TN 37235}} \maketitle

\begin{abstract}
A classification is given of Higgs-flavor models. In these models,
there are several Higgs doublets in an irreducible multiplet
$R_{\Phi}$ of a non-abelian symmetry $G_{\Phi}$, under which the
quarks and leptons do not transform (thus giving minimal
flavor-changing for the fermions). It is found that different
$G_{\Phi}$ and $R_{\Phi}$ lead to very distinctive spectra of the
extra Higgs doublets, including different numbers of ``sequential
Higgs" and of ``inert Higgs" that could play the role of dark
matter, different mass relations, and different patterns of
$SU(2)_L$-breaking splittings within the Higgs doublets.
\end{abstract}

\section{Introduction}

Just as there is a repetition of the quark and lepton ``families,"
it is plausible to suppose that there might be a repetition of Higgs
doublets \cite{Ajaib:2011my}. One theoretical difficulty with
additional Higgs doublets is that they could exacerbate the gauge
hierarchy problem, since the mass of each doublet might need to be
fine-tuned to have masses much less than the ``natural" Planck scale
or grand unification scale. A second difficulty is that a
multiplicity of Higgs doublets tends to lead to unrealistically
large flavor-changing neutral current (FCNC) processes. Here we
study the recently proposed idea of ``Higgs-flavor symmetry"
\cite{Barr:2011yq}. In models based on this idea, there are no
fine-tunings besides that of the Standard Model (SM) Higgs field,
and no FCNC effects aside from those caused by CKM mixing.

In these models, the ``extra" Higgs doublets and the Standard Model
(SM) Higgs doublet together form an irreducible multiplet $\Phi$ of
a non-abelian ``Higgs-flavor" group $G_{\Phi}$. Consequently, a
single tuning of the mass of the multiplet $\Phi$ (perhaps anthropic
\cite{anthropic}) is sufficient to make all the Higgs doublets
contained in it have low mass. The symmetry $G_{\Phi}$, in other
words, ties all the Higgs masses together.

Of course, the multiplet $\Phi$ must be split by $G_{\Phi}$-breaking
effects. It is assumed that the breaking of $G_{\Phi}$ occurs in a
sector of superheavy fields that are singlets under the SM group and
that the breaking is communicated to the SM fields (including
$\Phi$) by a ``messenger field" $\eta$ \cite{Barr:2011yq}.  The
vacuum expectation value (VEV) of $\eta$ is assumed to be near the
weak scale; but this does not require any fine-tuning, since the
{\it mass} of $\eta$ is superlarge. To make this concrete, suppose
that $G_{\Phi}$ is spontaneously broken by a fermion-antifermion
condensate $\langle \overline{\chi} \chi \rangle$ to which $\eta$
couples. Then the terms $y \overline{\chi} \chi \eta + \frac{1}{2}
M_{\eta}^2 \eta^2$ give $\langle \eta \rangle = y \langle
\overline{\chi} \chi \rangle/ M_{\eta}^2$. Even if $M_{\eta}^2$ is
superlarge, $\langle \eta \rangle$ can be of any scale without
fine-tuning, since the scale of $\langle \overline{\chi} \chi
\rangle$ is dynamically generated by some confining interaction, and
its value is determined by the scale at with that interaction
becomes strong, which can ``naturally" be anything. It should be
noted that if the symmetry $G_{\Phi}$ is local, the gauge bosons
associated with it get mass at the scale of the condensate $\langle
\overline{\chi} \chi \rangle$, which is of order $(M^2_{\eta}
\langle \eta \rangle)^{1/3} \gg M_W$. Therefore the FCNC effects
produced by their exchange are negligible. At low-energy, there is
effectively an approximate global $G_{\Phi}$ symmetry in the Higgs
sector, the pattern of breaking of which is determined by the
expectation value of the messenger field.

To avoid dangerous FCNC effects, one makes the crucial assumption
that the SM quarks and leptons are singlets under the group
$G_{\Phi}$. The quark and lepton masses thus must come from
higher-dimension effective operators. For example, the up quark
masses would come from operators of the form

\begin{equation}
Y^u_{mn} u^c_m u_n (\Phi^{\alpha} \eta^*_{\alpha})/M,
\end{equation}

\noindent and similarly for the down quarks and leptons, where $m,n$
are quark/lepton family indices, and $\alpha$ is a $G_{\Phi}$ index.
Note that for such a dimension-5 effective Yukawa operator to exist
the Higgs multiplet $\Phi$ and the messenger multiplet $\eta$ must
transform as the same irreducible representation $R_{\Phi}$ of
$G_{\Phi}$. The scale $M$ comes from integrating out fermions whose
mass is of order $\langle \eta \rangle$ (as was discussed in
\cite{Barr:2011yq}), so that the quark and lepton masses coming from
Eq. (1) need not be much suppressed compared to the weak scale.

From Eq. (1), one sees that only one linear combination of the Higgs
doublets (which will be called $\Phi_F$, where $F$ stands for
``fermions") couples to the quarks and leptons:

\begin{equation}
\Phi_F \propto \langle \eta^*_{\alpha} \rangle \Phi^{\alpha}.
\end{equation}

\noindent This satisfies the well-known conditions for ``natural
flavor conservation" \cite{nfc}. This linear combination is not
necessarily a mass eigenstate, but may be a mixture of the Standard
Model doublet (which we will denote $\Phi_{SM}$) and some of the
``extra" Higgs doublets. In that case, those extra doublets will
couple to quarks and leptons, but with Yukawa coupling matrices that
are proportional to those of $\Phi_{SM}$, so that no FCNC effects
result except through CKM mixing. (The idea of an extra Higgs
doublet whose Yukawa couplings to quarks and leptons are
proportional to those of the Standard Model Higgs has been proposed
recently by Pich and Tuz\'{o}n and further developed by Ser\^{o}dio
\cite{proportional}.)

One could imagine that there are several messenger fields; but then
there is a danger of excessive FCNC effects. For example, suppose
there were two messengers, $\eta$ and $\eta'$, which were in the
same representation of $G_{\Phi}$ as the Higgs multiplet $\Phi$.
Then two kinds of Yukawa term would be present for each type of
fermion; for example, for the up quarks one would have $Y^u_{mn}
u^c_m u_n (\Phi^{\alpha} \eta^*_{\alpha})/M$ and $Y^{\prime u}_{mn}
u^c_m u_n (\Phi^{\alpha} \eta^{\prime
*}_{\alpha})/M$, etc. This would mean that two different linear
combinations of the Higgs doublets would couple to quarks and
leptons with Yukawa matrices that were not, in general,
simultaneously diagonalizable. This would violate the condition for
natural flavor conservation \cite{nfc}. In this paper, therefore, we
will consider only the simplest case, namely: {\it there exists only
one messenger field, which is in the same irreducible representation
$R_{\Phi}$ of $G_{\Phi}$ as the Higgs multiplet} $\Phi$.

As was pointed out in \cite{Barr:2011yq}, and will be seen in the
cases worked out below, the framework we have just sketched leads to
models having several typical characteristics: (1) There exist one
or more Higgs doublets that are unable to decay to other Higgs
doublets or to fermions due to subgroups of $G_{\Phi}$ that are
unbroken by the messenger field VEV. The lightest components of such
doublets will be absolutely stable and play the role of dark matter.
(We will sometimes speak loosely of these doublets as ``stable Higgs
doublets," even though only their lightest components are stable.
The heavier components can decay into the lighter components plus
quarks or leptons by weak interactions.) (2) There exist in many
cases extra Higgs doublets that couple to quarks and leptons
proportionally to the Standard Model Higgs doublet (and therefore to
the masses of the fermions). These obviously are unstable. The
constants of proportionality are in some cases constrained by
$G_{\Phi}$. (3) There usually exist relationships coming from
$G_{\Phi}$ among the masses of the extra Higgs doublets and also
between these masses and the constants of proportionality mentioned
in (2). (4) There usually exist relationships coming from $G_{\Phi}$
among the $SU(2)_L$-breaking mass splittings within the Higgs
doublets.

\section{General Discussion and Summary of Results}

The breaking of the Higgs-family symmetry $G_{\Phi}$ is communicated
to the multiplet of Higgs doublets $\Phi$ by the messenger field
$\eta$, and in particular by terms in the Higgs potential that
couple $\Phi$ to $\eta$:

\begin{equation}
V_{HFB} \sim \Phi^{\dag} \cdot \Phi \; \eta^{\dag} \eta,
\end{equation}

\noindent where here and throughout the dot stands for the
contraction of electroweak indices to form an $SU(2)_L$ singlet. The
subscript ``HFB" stands for ``Higgs-flavor breaking." In some cases,
the group theory also allows terms of the form $\Phi^{\dag} \cdot
\Phi \eta$. Of course, there is always also a $G_{\Phi}$-invariant
mass term $M^2_0 \Phi^{\dag}_{\alpha} \cdot \Phi^{\alpha}$. When the
VEV of $\eta$ is substituted into $M_0^2 \Phi^{\dag} \cdot \Phi +
V_{HFB}$, one obtains the mass-squared matrix for the Higgs
doublets. It is assumed that $M_0^2$ is tuned so that the lightest
eigenstate has a negative mass-squared that is of order the weak
scale. This is the Standard Model Higgs doublet, which we will
denote $\Phi_{SM}$. The other eigenstates are assumed to have
positive mass-squared. These are the ``extra Higgs doublets," and
have no vacuum expectation values. (Technically, since they don't
obtain VEVs, one should not call them ``Higgs" fields. We shall,
nevertheless, do so for simplicity.)

One linear combination of the Higgs doublets couples to the Standard
Model quarks and leptons, namely that shown in Eq. (2). In some
cases, this $\Phi_F$ is a mass-squared eigenstate, in which case, it
must clearly be identified with $\Phi_{SM}$ if the quarks and
leptons are to obtain mass. Frequently however, as will be seen,
$\Phi_F$ is a linear combination of $\Phi_{SM}$ and one or more of
the ``extra Higgs doublets." Those extra Higgs would therefore
couple to the Standard Model quarks and leptons, and therefore be
able to decay directly into them. We will call these ``sequential
Higgs doublets." Their Yukawa coupling matrices are exactly
proportional to those of the Standard Model Higgs, as is clear from
Eq. (1). The constant of proportionality for each sequential Higgs
doublet is just given by the magnitude of its mixing with $\Phi_F$.
That is, if $\Phi_F = c_0 \Phi_{SM} + \sum_{\kappa} c_{\kappa}
\Phi_{\kappa}$, where $\Phi_{\kappa}$ stands for the ${\kappa}^{th}$
sequential Higgs doublet, and $c_0^2 + \sum_{\kappa} |c_{\kappa}|^2
=1$, then the Yukawa coupling matrix of the ${\kappa}^{th}$
sequential Higgs doublet is given by $(Y_{\kappa})_{mn} =
(c_{\kappa}/c_0)(Y_{SM})_{mn}$. We shall refer to the $c_{\kappa}$
as the ``mixing angles" of the sequential Higgs doublets.

In most cases, there are also Higgs-doublets that are mass-squared
eigenstates and do not mix with $\Phi_F$. These do not couple to the
quarks and leptons, and we shall call these ``inert Higgs doublets."
(Inert Higgs fields have been widely discussed as candidates for
dark matter \cite{inert}, though most of those models posit only one
inert Higgs field, whereas the ``Higgs-flavor" framework we are
discussing here typically leads to the existence of several inert
Higgs fields.) An inert Higgs doublet cannot decay directly into
quarks and leptons, but in some cases can decay into another inert
Higgs doublet plus some sequential Higgs that can then decay into
quarks and leptons. Thus, some inert Higgs doublets are unstable. As
will be seen, however, some of the inert Higgs doublets are
absolutely stable. More precisely, their lightest components are
absolutely stable, since these doublets are split when the weak
gauge group $SU(2)_L$ breaks, and their heavier components can beta
decay into the lightest one. We shall nevertheless sometimes for
simplicity refer to these as ``stable Higgs doublets." We shall
assume that the $SU(2)_L$ breaking is such that the stable
components of the ``stable Higgs doublets" are neutral rather than
charged, simply because there are very stringent experimental limits
on stable charged particles. This can always be ensured by a choice
of signs of certain quartic self couplings of the Higgs doublets.

What makes the ``stable Higgs doublets" stable, as will be seen, are
subgroups of the Higgs-flavor group that are left unbroken by the
messenger field expectation value. These subgroups may be discrete
or continuous.

The $SU(2)_L$-breaking splittings within the extra Higgs doublets
are produced by the quartic part of the Higgs potential, which has
the form

\begin{equation}
V_4 \sim \Phi^{\dag} \cdot \Phi \; \; \Phi^{\dag} \cdot \Phi.
\end{equation}

\noindent When expanded out, $V_4$ contains terms that contain two
powers of $\Phi_{SM}$ and two powers of ``extra Higgs doublets."
(There are, of course, also terms that are higher order in the extra
Higgs doublets, but these don't contribute to splitting those
doublets.) In some cases, an extra Higgs doublet only has a
splitting between its charged and (complex) neutral components. We
shall call this a ``$CN$" splitting. In other cases, an extra Higgs
doublet also has a splitting between the pseudoscalar and
scalar part of its neutral component. We shall call this a ``$CPS$"
splitting.

We shall consider all cases involving continuous Higgs-flavor groups
(though they may also have discrete factors) that have six or fewer
Higgs doublets. Each case is characterized by a choice of the group
$G_{\Phi}$ and the irreducuble representation $R_{\Phi}$ to which
(by assumption) both $\Phi$ and $\eta$ belong. Nearly every case gives a
very distinctive spectrum of extra Higgs doublets.

In Table I is displayed, for each case considered, the number of extra
Higgs doublets of the following types: (a) sequential Higgs
doublets, (b) inert Higgs doublets that are $CN$-split, and (c)
inert Higgs doublets that are $CPS$-split. If a number $N >1$
appears as an entry, it refers $N$ degenerate Higgs doublets. So
$(1,1,1,1)$ means that there are four Higgs doublets that are not
degenerate, $(2,2)$ means that there are two doublets of one mass
and two of a different mass, and $4$ means that there are four Higgs
doublets of the same mass, etc. If there is no superscript on a
number, it means that the lightest component(s) of those Higgs
doublets are absolutely stable. For a $CN$-split Higgs doublet, the
lightest component is a complex neutral field. For a $CPS$-split
Higgs doublet, it is a real field. So, for example, if there is a 4
in the $CN$ column, it means that there are 4 stable complex neutral
fields that are degenerate in mass. If there is a 4 in the CPS
column, it means that there are 4 stable real fields that are
degenerate. A superscript zero means that the corresponding Higgs
doublet does {\it not} have a stable component. A superscript
asterisk means that whether it does decay or not depends on the values of
certain parameters in the model.

Making  use of the isomorphisms $SO(3) \sim SU(2)$, $SO(5) \sim Sp(4)$,
$SO(4) \sim SU(2)\times SU(2)$, and $SO(6) \sim SU(4)$, one can see that all
cases of representations of dimension less than or equal to 6 have been included
in Table 1.

\newpage

\noindent {\large\bf Table I:} The number of extra Higgs doublets of
various types in different Higgs-flavor schemes. The notation is
defined in the text.

\vspace{0.2cm}

\begin{tabular}{|l|l||l|l|l|}
\hline {\bf $G_{\Phi}$} & $R_{\Phi}$ & Sequential & Inert Higgs & Inert Higgs\\
& & Higgs & $CN$-split & $CPS$-split \\
\hline $SU(2)$ & 2 & - & 1 & - \\
\hline $SU(3)$ & 3 & - & 2 & - \\
\hline $SO(3)$ & $3_R$ & - & (1,1) & - \\
\hline $SO(3)$ & $3_R$ ($m_3 =0)$ & - & - & 2 \\
\hline $SO(3)$ & $3_C$ & 1 & - & 1 \\
\hline $SU(4)$ & 4 & - & 3 & - \\
\hline $SO(4)$ & $4_R$ & - & - & 3 \\
\hline $SO(4)$ & $4_C$ & 1 & - & $(1,1)$ \\
\hline $SU(2)$ & 4 & $(1,1,1)$ & - & - \\
\hline $SU(2)^2 \times S_2$ & $(2,1) + (1,2)$ & 1 & - &
$(1,1^0)$ \\
\hline $SO(5)$ & 4 & 1 & 2 & - \\
\hline $SU(5)$ & 5 & - & 4 & - \\
\hline $SO(5)$ & $5_R$ & - & - & 4 \\
\hline $SO(5)$ & $5_C$ & 1 & - & 3 \\
\hline $SO(3)$ & $5_R$ & 1 & - & $(1,1,1^*)$ \\
\hline $SO(3)$ & $5_C$ & $(1,1,1,1)$ & - & - \\
\hline $SU(6)$ & 6 & - & 5 & - \\
\hline $SO(6)$ & $6_R$ & - & - & 5 \\
\hline $SO(6)$ & $6_C$ & 1 & - & 4 \\
\hline $SU(3)$ & 6 & $(1,1)$ & - & $(1,1,1^*)$ \\
\hline $SU(2)$ & 6 & $(1,1,1,1,1)$ & - & - \\
\hline $SU(3)^2 \times S_2$ & $(3,1) + (1,3)$ & 1 &
$(2,2)$ & - \\
\hline $SO(3)^2 \times S_2$ & $[(3,1) + (1,3)]_R$ & 1 & (1,1,1,1)
& - \\
\hline $SO(3)^2 \times S_2$ & $[(3,1) + (1,3)]_C$ &
$(1,1,1)$ & - & $(1,1)$ \\
\hline $SU(3) \times SU(2)$ & $(3,2)$ & 1 & $(1,1^*)$ &
$(1^*, 1^0)$ \\
\hline $SU(2)^2$ & $(3,2)$ & $(1,1,1,1,1)$ & - & - \\
\hline $SU(2)^3 \times S_3$ & $(2,1,1)$ & $(1,1)$ & $(1,1,1)$ & - \\
& $+ (1,2,1) + (1,1,2)$ & & & \\
\hline
\end{tabular}

\vspace{0.5cm }

The Higgs-flavor symmetry highly restricts the form of the Higgs
potential, and thus gives relations among the masses and mixings of
the Higgs doublets. We analyse the Higgs mass spectrum in two
stages, as explained above. First, there is the spectrum in the
limit where $SU(2)_L$ breaking is neglected. This is found from the
mass-squared matrix that comes from the terms $M_0^2 \Phi^{\dag}
\cdot \Phi + V_{HFB}$. The measureable quantities of interest are
the values of the masses of the extra Higgs doublets and the values
of the mixing angles of the sequential Higgs doublets with $\Phi_F$
(which completely determine their coupings to the Standard Model
quarks and leptons). Call the number of extra Higgs doublets
$N_{EH}$, the number of {\it distinct} masses of the extra Higgs
doublets (i.e. treating Higgs doublets that are degenerate with each
other as having one mass) $N_M$, and the number of sequential Higgs
doublets (and thus of mixing angles) $N_{\theta}$. These $N_M +
N_{\theta}$ quantities are determined by the values of two kinds of
model parameters: (i) coefficients in the potential $V_{HFB}$ that
contribute to splitting among Higgs doublets, and (ii) the number of
independent physical quantities in the VEV of the messenger field.
Call the total number of these two kinds of model parameters
$N_{par}$. Then there will be $N_{rel} = \max(N_M + N_{\theta} -
N_{par},0)$ relations among the masses and mixings of the Higgs
doublets.

The second stage of analysis is to find the $SU(2)_L$-breaking
splittings {\it within} Higgs doublets. Each Higgs doublet (or
degenerate set of doublets) that is $CN$-split has one measureable
splitting, whereas each that is $CPS$-split has two measureable
splittings. Call the total number of measureable splittings
$N_{split}$. These depend on the number $N_4$ of coefficients in
$V_4$ that contribute to such splittings. There will therefore be a
further $N'_{rel}$ relations where $N'_{rel} = N_{split} - N_4$
(unless $N_M + N_{\theta} - N_{par}$ was negative, in which case
$N'_{rel}$ is reduced by that amount).

In Table II, we display these numbers for all the cases where
$G_{\Phi}$ contains continuous groups and that have six or fewer
Higgs doublets.

\newpage

\noindent {\large\bf Table II:} This shows the number of distinct
masses and mixing angles, the number of model parameters on which
they depend, and the number of relations predicted, for each model.
The quantities are defined in the text.

\vspace{0.2cm}

\begin{tabular}{|l|l||l||l|l|l|l||l|l|l|}
\hline {\bf $G_{\Phi}$} & $R_{\Phi}$ & $N_{EH}$ & $N_M$ &
$N_{\theta}$ & $N_{par}$ & $N_{rel}$ & $N_{split}$ & $N_4$ &
$N'_{rel}$ \\
\hline $SU(2)$ & 2 & 1 & 1 & 0 & 1 & - & 1 & 1 & - \\
\hline $SU(3)$ & 3 & 2 & 1 & 0 & 1 & - & 1 & 1 & - \\
\hline $SO(3)$ & $3_R$ & 2 & 2 & 0 & 2 & - & 2 & 2 & - \\
\hline $SO(3)$ & $3_R$ ($m_3 =0)$ & 2 & 1 & 0 & 1 & - & 2 & 2 & - \\
\hline $SO(3)$ & $3_C$ & 2 & 2 & 1 & 3 & - & 4 & 2 & 2 \\
\hline $SU(4)$ & 4 & 3 & 1 & 0 & 1 & - & 1 & 1 & - \\
\hline $SO(4)$ & $4_R$ & 3 & 1 & 0 & 1 & - & 2 & 2 & - \\
\hline $SO(4)$ & $4_C$ & 3 & 3 & 1 & 4 & - & 4 & 3 & 1 \\
\hline $SU(2)$ & 4 & 3 & 3 & 3 & 6 & - & 6 & 3 & 3 \\
\hline $SU(2)^2 \times S_2$ & $(2,1) + (1,2)$ & 3 & 3 & 1 & 5 & - & 4 & 4 & - \\
\hline $SO(5)$ & 4 & 3 & 2 & 1 & 3 & - & 3 & 3 & - \\
\hline $SU(5)$ & 5 & 4 & 1 & 0 & 1 & - & 1 & 1 & - \\
\hline $SO(5)$ & $5_R$ & 4 & 1 & 0 & 1 & - & 2 & 2 & - \\
\hline $SO(5)$ & $5_C$ & 4 & 2 & 1 & 3 & - & 4 & 2 & 2 \\
\hline $SO(3)$ & $5_R$ & 4 & 4 & 1 & 4 & 1 & 8 & 4 & 4 \\
\hline $SO(3)$ & $5_C$ & 4 & 4 & 4 & 9 & - & 8 & 4 & 3\\
\hline $SU(6)$ & 6 & 5 & 1 & 0 & 1 & - & 1 & 1 & - \\
\hline $SO(6)$ & $6_R$ & 5 & 1 & 0 & 1 & - & 2 & 2 & - \\
\hline $SO(6)$ & $6_C$ & 5 & 2 & 1 & 3 & - & 4 & 2 & 2 \\
\hline $SU(3)$ & 6 & 5 & 5 & 2 & 4 & 3 & 10 & 2 & 8 \\
\hline $SU(2)$ & 6 & 5 & 5 & 5 & 12 & - & 10 & 5 & 3 \\
\hline $SU(3)^2 \times S_2$ & $(3,1) + (1,3)$ & 5 & 3
& 1 & 4 & - & 4 & 4 & - \\
\hline $SO(3)^2 \times S_2$ & $[(3,1) + (1,3)]_R$ &
5 & 5 & 1 & 6 & - & 6 & 6 & 4 \\
\hline $SO(3)^2 \times S_2$ & $[(3,1) + (1,3)]_C$ & 5 & 5 & 3 & 8 & - & 10 & 6 & 4 \\
\hline $SU(3) \times SU(2)$ & $(3,2)$ & 5 & 5 & 1 & 4 & 2 & 8 & 3 & 5 \\
\hline $SU(2)^2$ & $(3,2)$ & 5 & 5 & 5 & 12 & - & 10 & 5 & 3 \\
\hline $SU(2)^3 \times S_3$ & $(2,1,1)$ & 5 & 5 & 2 & 5 & 2 & 7 & 3 & 4 \\
& $+ (1,2,1)$ & & & & & & & & \\
& $+ (1,1,2)$ & & & & & & & &  \\ \hline
\end{tabular}

\vspace{0.5cm }

\section{Analysis of Cases}

We now analyse in some detail cases where the number of Higgs
doublets is six or fewer. Each case is defined by the Higgs-flavor
group $G_{\Phi}$ and the irreducible representation $R_{\Phi}$ to
which (by our assumption) both $\Phi$ and $\eta$ belong.

\subsection{$G_{\Phi} = SU(N)$ with $R_{\Phi} = N$.}

\noindent {\bf (a)} If the messenger field is in the fundamental
representation of $SU(N)$, then its VEV can, without loss of generality, be
brought to the form

\begin{equation}
\langle \eta^{\alpha} \rangle = \left( \begin{array}{c} \eta \\ 0 \\
.
\\ . \\ 0 \end{array} \right),
\end{equation}

\noindent where $\eta$ is real. Then

\begin{equation}
\Phi_F \propto \langle \eta^*_{\alpha} \rangle \Phi^{\alpha}
\Rightarrow \Phi_F = \Phi^1.
\end{equation}

\noindent The breaking of $SU(N)$ is communicated to the Higgs
doublets through terms of the form $\Phi^{\dag} \cdot \Phi
\eta^{\dag} \eta$. The $SU(N)$ indices of the product $\eta^{\dag}
\eta$ can be contracted in two ways, corresponding to the
decomposition $\overline{{\bf N}} \times {\bf N} = {\bf 1} + {\bf
Adj}$. There are therefore two independent terms in $V_{HFB}$:

\begin{equation}
V_{HFB} = \sigma_1 \; \Phi^{\dag}_{\alpha} \cdot \Phi^{\alpha} \;
\eta^*_{\beta} \eta^{\beta} + \sigma_2 \; \Phi^{\dag}_{\alpha} \cdot
\Phi^{\beta} \; \eta^*_{\beta} \eta^{\alpha},
\end{equation}

\noindent where the $\sigma_i$ are real. Substituting in Eq. (7) the
VEV of $\eta^{\alpha}$ given in Eq. (5), one has

\begin{equation}
\left( \begin{array}{cccc} \Phi^1 & \Phi^2 & .. & \Phi^N
\end{array} \right)^{\dag} \cdot \left( \begin{array}{cccc} m^2 +
\sigma_2 \eta^2 & 0 & . & 0 \\ 0 & m^2 & . & 0 \\ . & . & . & . \\
0 & 0 & . & m^2 \end{array} \right) \left( \begin{array}{c} \Phi^1
\\ \Phi^2 \\ : \\ \Phi^N \end{array} \right),
\end{equation}

\noindent where $m^2 \equiv M_0^2 + \sigma_1 \eta^2$. Since, only
$\Phi^1$ couples to quarks and leptons, it must be the Standard
Model Higgs doublet $\Phi_{SM}$, and thus the lightest doublet. One
must therefore assume that $\sigma_2 \eta^2 < 0$ and that $M_0^2$
(and thus $m^2$) is fine-tuned (presumably anthropically
\cite{anthropic}) so that $m^2 + \sigma_2 \eta^2 = - |\mu|^2$, where
$|\mu|^2$ is of order (100 GeV)$^2$. All the other Higgs doublets
$\Phi^{\alpha}$, $\alpha = 2, ..., N$, are degenerate and have
mass-squared given by $m^2 = - |\mu|^2 + |\sigma_2| \eta^2$, which
we take to be positive and large enough that these extra Higgs
doublets have so far evaded detection
--- how large depends on the size of the messenger VEV $\eta$, which
can ``naturally" be anything, as noted before.

Given the form of the messenger field VEV in Eq. (5), the Higgs
sector of the low-energy effective theory has a residual $SU(N-1)$
global symmetry. Under this symmetry, the extra $N-1$ Higgs doublets
form an $(N-1)$-plet, whereas all the Standard Model fields are
singlets. This symmetry therefore prevents the decays of the extra
Higgs doublets to Standard Model fields. There are, of course,
$SU(2)_L$-breaking splittings within the extra Higgs doublets.
Consequently, one component of an extra Higgs doublet can beta decay
to a lighter component plus an quark-antiquark or lepton-antilepton
pair. But the lightest component of each of the $N-1$ extra Higgs
doublets is absolutely stable. And these $N-1$ stable particles are
all degenerate due to the residual global $SU(N-1)$.

We now turn to the $SU(2)_L$-breaking splittings within the
doublets. These come from the quartic terms in the Higgs potential
that are of the form $\Phi^{\dag} \cdot \Phi \Phi^{\dag} \cdot
\Phi$. The $SU(N)$ indices of the product $\Phi^{\dag} \cdot \Phi$
can be contracted in two ways, corresponding to the decomposition
$\overline{{\bf N}} \times {\bf N} = {\bf 1} + {\bf Adj}$. There are
therefore two terms in $V_4$:

\begin{equation}
V_4 = \lambda_1 \Phi^{\dag}_{\alpha} \cdot \Phi^{\alpha} \;\;
\Phi^{\dag}_{\beta} \cdot \Phi^{\beta} + \lambda_2
\Phi^{\dag}_{\alpha} \cdot \Phi^{\beta} \;\; \Phi^{\dag}_{\beta}
\cdot \Phi^{\alpha},
\end{equation}

\noindent where the $\lambda_i$ are real. Since $\Phi^1 =
\Phi_{SM}$, it has a negative mass-squared and its neutral component
has a VEV, $\langle (\Phi^1)^0 \rangle = v/\sqrt{2}$, whereas all
the other Higgs doublets $\Phi^{\alpha}$, $\alpha = 2, .., N$ have
positive mass-squared and vanishing VEVs. Substituting this
expectation value into Eq. (9), it is easy to see that there is a
mass term $\lambda_2 v^2 |(\Phi^{\beta})^0|^2$, for $\beta = 2, ...,
N$, which splits the charged components of the extra Higgs doublets
from their neutral components, by the same amount for every $\beta$.
Note that there is no splitting between the scalar and pseudoscalar
components of these doublets, i.e. they are $CN$-split. Since the
dark matter must be neutral, one concludes that $\lambda_2 < 0$. The
dark matter fields, therefore, consist of $N-1$ degenerate neutral
complex fields.

\subsection{$G_{\Phi} = SO(N)$ with $R_{\Phi} = N$, $\eta$ = real.}

\subsubsection{ $N > 3.$}

The analysis of this case is quite similar to the previous one. We
will use latin letters to denote $SO(N)$ indices. Without loss of
generality, the VEV of the messenger field can be brought to the
form

\begin{equation}
\langle \eta^i \rangle = \left( \begin{array}{c} \eta \\ 0 \\
.
\\ . \\ 0 \end{array} \right),
\end{equation}

\noindent The Higgs doublet that couples to the fermions is
therefore again $\Phi_F = \Phi^1$. Since $\eta$ is now a real field,
Eq. (3) becomes $V_{HFB} \sim \Phi^{\dag} \cdot \Phi \eta \eta$. The
product $\eta \eta$ must now be in the {\it symmetric} $SO(N)$
product $({\bf N} \times {\bf N})_S = {\bf 1} + {\bf
(\frac{N(N+1)}{2} -1)}$. There are therefore two terms given by

\begin{equation}
V_{HFB} = \sigma_1 \; \Phi^{i \dag} \cdot \Phi^i \; \eta^j \eta^j +
\sigma_2 \; \Phi^{i \dag} \cdot \Phi^j \; \eta^i \eta^j,
\end{equation}

\noindent Upon substituting the VEV of the messenger field, one
obtains again the mass matrix given in Eq. (8). As in the previous
case, one can make $\Phi^1$ the lightest
Higgs doublet by choosing $\sigma_2 < 0$, and one can make it have negative
mass-squared by tuning $M_0^2$, so that $\Phi_{SM} = \Phi_F = \Phi^1$. The VEV of the
messenger field leaves unbroken a global $SO(N-1)$, so that all the
extra Higgs doublets are, as before, degenerate. Moreover, this
unbroken $SO(N-1)$ causes the lightest components of all the extra
doublets to be stable.

The main difference with the previous case lies in the form of
$V_4$. Since $\Phi^{\dag} \cdot \Phi$ is in the product (not
required to be symmetric) ${\bf N} \times {\bf N} = {\bf 1} + {\bf +
{\bf \frac{N(N-1)}{2}} + (\frac{N(N+1)}{2} -1)}$, there are three
independent terms in $V_4$, which can be written

\begin{equation}
V_4 = \lambda_1 \Phi^{i \dag} \cdot \Phi^i \;\; \Phi^{j \dag} \cdot
\Phi^j + \lambda_2 \Phi^{i \dag} \cdot \Phi^j \;\; \Phi^{j \dag}
\cdot \Phi^i + \lambda_3 \Phi^{i \dag} \cdot \Phi^j \;\; \Phi^{i
\dag} \cdot \Phi^j.
\end{equation}

\noindent (One might imagine that $N=4$ is a special case here,
since the existence of a rank-4 epsilon symbol allows one to write a
term $\Phi^{i \dag} \cdot \Phi^j \;\; \Phi^{k \dag} \cdot
\Phi^{\ell} \epsilon^{ijk \ell}$. However, this has no effect on the
$SU(2)_L$-breaking splittings, since only $\Phi^1$ has a non-zero
VEV.)

The second and third terms in Eq. (12) give the following
$SU(2)_L$-breaking contributions to the mass-squared of the Higgs
doublets: $\lambda_2 v^2 |(\Phi^j)^0|^2 + \lambda_3 v^2
[((\Phi^j)^0)^2 + ((\Phi^j)^{0*})^2]$. Note that while the
$\lambda_2$ term only splits the charged from the neutral
components, as in the $SU(N)$ case, the $\lambda_3$ term now also
splits the scalar from the pseudoscalar in the neutral component. So
the extra Higgs doublets are $CPS$-split.

The dark matter would be either the neutral scalar or neutral
pseudoscalar components of the $\Phi^i$, $i = 2,...,N$, depending on
which is lighter. Thus, the dark matter fields consist of $N-1$
degenerate real neutral fields.

\subsubsection{$N=3$.}

The case of $N=3$ is special, because the existence of the rank-3
epsilon symbol $\epsilon^{ijk}$ allows an additional term in
$V_{HFB}$:

\begin{equation}
V_{HFB} = \sigma_1 \; \Phi^{i \dag} \cdot \Phi^i \; \eta^j \eta^j +
\sigma_2 \; \Phi^{i \dag} \cdot \Phi^j \; \eta^i \eta^j + i m_3
\Phi^{i \dag} \cdot \Phi^j \; \eta^k \; \epsilon^{ijk}.
\end{equation}

\noindent This gives a mass-squared matrix

\begin{equation}
\left( \begin{array}{ccc} \Phi^1 & \Phi^2 & \Phi^3
\end{array} \right)^{\dag} \cdot \left( \begin{array}{ccc} m^2 +
\sigma_2 \eta^2 & 0 & 0  \\ 0 & m^2 & i m_3 \eta \\ 0 & -i m_3 \eta
& m^2 \end{array} \right) \left( \begin{array}{c} \Phi^1
\\ \Phi^2 \\ \Phi^3 \end{array} \right),
\end{equation}

\noindent where $m^2 = M_0^2 + \sigma_1 \eta^2$. The two extra Higgs
doublets are therefore $\Phi_{\pm} = \frac{1}{\sqrt{2}} (\Phi^2 \pm
i \Phi^3)$, with mass-squared given by $m_{\pm} = m^2 \pm m_3 \eta$.
The VEV of the messenger field breaks the global $SO(3)$ down to
$SO(2) = U(1)$, which, being an abelian group, has only singlet
representations, and therefore doesn't cause any degeneracy among
the Higgs doublets. Under this $U(1)$ (which is a rotation in the 23
plane), $\Phi_{\pm} \rightarrow e^{\pm i \theta} \Phi_{\pm}$. This
symmetry prevents either $\Phi_{\pm}$ from decaying entirely into
Standard Model particles, which are all neutral under it. But it
does allow decays such as $\Phi_{\pm} \rightarrow \Phi^*_{\mp} +
...$, which can happen through quartic scalar couplings. Therefore,
only the lighter of the two extra Higgs doublets $\Phi_{\pm}$ has
stable components.

The quartic couplings are still described by Eq. (12); but now these
lead to a more complicated spectrum. Each of the two extra Higgs
doublets has three types of components: the charged component and
the real and imaginary parts of the neutral components, which will
be denoted by the subscripts $C$, $R$, and $I$, respectively. One
ends up with the following $SU(2)_L$-breaking contributions to the
mass-squared matrices of the extra Higgs doublets:

\begin{equation}
\begin{array}{l}
\left( \phi_{+C}, \phi_{-C} \right)^* \left[ \begin{array}{cc}  m_3
\eta & 0 \\ 0 & - m_3 \eta \end{array} \right] \left(
\begin{array}{c} \phi_{+C} \\ \phi_{-C} \end{array} \right)
\\ \\
\left( \phi_{+R}, \phi_{-R}, \phi_{+I}, \phi_{-I} \right) \left[
\begin{array}{cccc} m_3 \eta + \lambda_2 v^2 & \lambda_3 v^2 & 0 & 0
\\ \lambda_3 v^2 & -m_3 \eta + \lambda_2 v^2 & 0 & 0 \\
0 & 0 & m_3 \eta + \lambda_2 v^2 & - \lambda_3 v^2
\\ 0 & 0 & - \lambda_3 v^2 & -m_3 \eta + \lambda_2 v^2
\end{array}
\right] \left( \begin{array}{c} \phi_{+R}
\\ \phi_{-R} \\ \phi_{+I} \\ \phi_{-I} \end{array} \right).
\end{array}
\end{equation}

\noindent This gives the following masses:

\begin{equation}
\begin{array}{ll}
M^2(\phi_{+C}) & = M_0^2 + m_3 \eta, \\
M^2(\phi_{+R,I}) & = M_0^2 + \sqrt{(m_3 \eta)^2 + (\lambda_3 v^2)^2}
+
\lambda_2 v^2, \\
M^2(\phi_{-C}) & = M_0^2 - m_3 \eta, \\
M^2(\phi_{-R,I}) & = M_0^2 - \sqrt{(m_3 \eta)^2 + (\lambda_3 v^2)^2}
+ \lambda_2 v^2.
\end{array}
\end{equation}

\noindent One sees that the doublets $\Phi_+$ and $\Phi_-$ are both
$CN$-split. Moreover, the two splittings are determined by the two
parameters $\lambda_2$ and $\lambda_3$. So, there are no relations.
Because the dark matter must be neutral, the only stable particle is
the (complex) neutral component of the lighter of $\Phi_+$ and
$\Phi_-$. That is, the dark matter field consists of one complex
neutral field.

\subsection{$G_{\Phi} = SO(N)$ with $R_{\Phi} = N$, $\eta$ = complex.}

\subsubsection{$N \neq 4$.}

Here we assume that $\eta$ is a complex field in the $N$
representation of $SO(N)$. Besides the $SU(N)$ Higgs-flavor
symmetry, there is assumed to be a $U(1)$ (global or local) under
which $\eta$ is charged. $\Phi$ must have the same $U(1)$ charge so
that effective Yukawa terms for the quarks and leptons can be
written down of the form $Y_{mn} \psi_m \psi_n (\Phi \eta^*)$. (Note
that if $\eta$ were not charged under a $U(1)$, then two types of
effective Yukawa term could be written down, involving $\Phi \eta$
and $\Phi \eta^*$, whose Yukawa coupling matrices would not be
related by any symmetry, thus leading in general to excessive FCNC
effects.)

If $\eta$ is a complex field, then its VEV cannot be brought to the
form in Eq.(10) by $SO(N)$ rotations. However, it can be shown that
by a combination of $SO(N)$ rotation and rephasing of $\eta$ the VEV
can be brought to the following form without any loss of generality:

\begin{equation}
\langle \eta^i \rangle = \left( \begin{array}{c} \eta_1 \\ i \eta_2
\\ 0 \\.
\\ . \\ 0 \end{array} \right),
\end{equation}

\noindent where $\eta_1$ and $\eta_2$ are real. Therefore, the Higgs
doublet that couples to the fermions is $\Phi_F \propto \eta_1
\Phi^1 - i \eta_2 \Phi^2$.

In the terms of $V_{HFB}$, which are of the form $\Phi^{\dag} \cdot
\Phi \eta^{\dag} \eta$, the product $\eta^{\dag} \eta$ can be
contracted in three ways, corresponding to the decomposition ${\bf
N} \times {\bf N} = {\bf 1} +  {\bf \frac{N(N-1)}{2}} + {\bf
(\frac{N(N+1)}{2} -1)}$. Thus,

\begin{equation}
V_{HFB} = \sigma_1 \; \Phi^{i \dag} \cdot \Phi^i \; \eta^{j*} \eta^j
+ \sigma_2 \; \Phi^{i \dag} \cdot \Phi^j \; \eta^{j*} \eta^i +
\sigma_3 \Phi^{i \dag} \cdot \Phi^j \; \eta^{i*} \eta^j.
\end{equation}

\noindent This gives the mass matrix

\begin{equation}
\left( \begin{array}{cccc} \Phi^1 & \Phi^2 & .. & \Phi^N
\end{array} \right)^{\dag} \cdot \left( \begin{array}{cccc} m^2 +
(\sigma_2 + \sigma_3)\eta_1^2 & -i(\sigma_2 - \sigma_3) \eta_1
\eta_2 & . & 0 \\ i (\sigma_2 - \sigma_3) \eta_1 \eta_2 & m^2 +
(\sigma_3 + \sigma_2) \eta_2^2
& . & 0 \\ . & . & . & . \\
0 & 0 & . & m^2 \end{array} \right) \left( \begin{array}{c} \Phi^1
\\ \Phi^2 \\ : \\ \Phi^N \end{array} \right),
\end{equation}

\noindent where $m^2 \equiv M_0^2 + \sigma_1 (\eta_1^2+ \eta_2^2)$.
Therefore, the Higgs doublets $\Phi^1$ and $\Phi^2$ mix, and the
eigenstates of Eq. (19) are $\Phi \equiv \cos \theta \Phi^1 + i \sin
\theta \Phi^2$, and $\Phi' \equiv - \sin \theta \Phi^1 + i \cos
\theta \Phi^2$. The two distinct masses of the extra Higgs doublets
(that of $\Phi'$ and that of $\Phi^i$, $i = 3,...,N$) and the mixing
angle $\theta$ are determined by the three parameters, $\sigma_2
\eta_1^2$, $\sigma_3/\sigma_2$, and $\eta_2/\eta_1$. So, no relation
exists among them.

The fields $\Phi^i$, $i = 3, ..., N$ are degenerate due to an
$SO(N-2)$ that is left unbroken by the messenger VEV. They are
prevented by this symmetry from decaying entirely into Standard
Model fields, and so the lightest components of these Higgs doublets
are stable. Both $\Phi$ and $\Phi'$ couple to quarks and leptons,
and are therefore unstable. The lighter of these (which we can take
to be $\Phi$) must be the Standard Model Higgs doublet $\Phi_{SM}$,
whose mass-squared is negative (by the tuning of $M_0^2$).

The $SU(2)_L$-breaking splittings of the extra Higgs doublets,
$\Phi'$ and $\Phi^1$, $i=3, ..., N$, are determined by the quartic
terms, which have the same form as shown in Eq. (12). As in the case
where $\eta$ is real (discussed in section 3.2.1), all the extra
Higgs doublets are $CPS$-split. These splittings are the same for
the $\Phi^i$, $i = 3, ..., N$, by the residual $SO(N-2)$. It can be
shown that these splittings are in general different for the
$\Phi'$, however. There are four distinct $SU(2)_L$ splittings,
therefore, which are determined by the two parameters $\lambda_2$
and $\lambda_3$, giving two relations.

The dark matter consists of $N-2$ degenerate real neutral fields,
which are either the neutral scalar or neutral pseudoscalar
components of the $\Phi^i$, $i=3,...,N$, depending on which are
lighter.

\subsubsection{$N=4$.}

For $N=4$, additional terms are made possible by the existence of
the rank-4 epsilon symbol $\epsilon^{ijk \ell}$. In $V_{HFB}$ there
is an additional term besides those shown in Eq. (18), namely
$\sigma_4 \Phi^{i \dag} \cdot \Phi^j \; \eta^{k*} \eta^{\ell}
\epsilon^{ijk \ell}$. (One might think that there are only three
terms in $V_{HFB}$, since the product $\eta^{\dag} \eta$ must be in
${\bf 4} \times {\bf 4} = {\bf 1} + {\bf 6} + {\bf 9}$. However,
there are two ways to contract ${\bf 6} \times {\bf 6}$ to make a
singlet: $T^{[ij]}T^{[ij]}$ and $T^{[ij]} T^{[k \ell]} \epsilon^{ijk
\ell}$.)

The effect of the extra term is to mix $\Phi^3$ and $\Phi^4$, so
that the eigenstates of the mass-squared matrix produced by
$V_{HFB}$ are $\Phi \equiv \cos \theta \Phi^1 + i \sin \theta
\Phi^2$, $\Phi' \equiv - \sin \theta \Phi^1 + i \cos \theta \Phi^2$,
$\Phi_+ = \frac{1}{\sqrt{2}} (\Phi^3 + i \Phi^4)$, and $\Phi_- =
\frac{1}{\sqrt{2}} (\Phi^3 - i \Phi^4)$. The masses of the three
extra Higgs doublets and the mixing angle $\theta$ are determined by
four parameters, $\sigma_2 \eta_1^2$, $\sigma_3/\sigma_2$,
$\sigma_4/\sigma_2$, and $\eta_2/\eta_1$; so (as in the $N \neq 4$
case) no relation exists among them.

Under the $SO(2)$ left unbroken by the messenger VEV, which is a
rotation in the 34 plane, one has $\Phi_{\pm} \rightarrow e^{\pm i
\theta} \Phi_{\pm}$. This symmetry prevents either $\Phi_{\pm}$ from
decaying entirely into Standard Model particles. But it does allow
decays such as $\Phi_{\pm} \rightarrow \Phi^*_{\mp} + ...$, which
can happen through quartic scalar couplings. Therefore, only the
lighter of the two extra Higgs doublets $\Phi_{\pm}$ has stable
components.

The $SU(2)_L$-breaking splittings within the extra Higgs doublets
are produced by $V_4$, which has, for $N=4$, an extra term besides
those shown in Eq. (12), namely $\lambda_4 \Phi^{i \dag} \cdot
\Phi^j \; \Phi^{k \dag} \cdot \Phi^{\ell} \epsilon^{ijk \ell}$. One
finds that $\Phi_{\pm}$ are both $CN$-split. (Note that this is
different from the situation for $N \neq 4$. Also note that this
means that the stable extra Higgs field is a single {\it complex}
scalar.) This can be seen from the $4 \times 4$ mass matrix of these
neutrals, which has the form

\begin{equation}
\left( \phi_{+R}, \phi_{-R}, \phi_{+I}, \phi_{-I} \right) \left[
\begin{array}{cccc} \overline{m}^2 + \sigma_4 \eta_1 \eta_2 & \overline{\lambda}_3 v^2 & 0 & 0
\\ \overline{\lambda}_3 v^2 & \overline{m}^2 - \sigma_4 \eta_1 \eta_2 & 0 & 0 \\
0 & 0 & \overline{m}^2 + \sigma_4 \eta_1 \eta_2 & -
\overline{\lambda}_3 v^2
\\ 0 & 0 & - \overline{\lambda}_3 v^2 & \overline{m}^2 \eta - \sigma_4 \eta_1 \eta_2
\end{array}
\right] \left( \begin{array}{c} \phi_{+R}
\\ \phi_{-R} \\ \phi_{+I} \\ \phi_{-I} \end{array} \right),
\end{equation}

\noindent where $\overline{m}^2 \equiv M_0^2 + \sigma_1 (\eta_1^2 +
\eta_2^2) + (\lambda_1 + \lambda_2) v^2$, and $\overline{\lambda}_3
\equiv \lambda_3 (\cos^2 \theta - \sin^2 \theta)$. This has just two
distinct eigenvalues: $\overline{m}^2 \pm \sqrt{(\sigma_4 \eta_1
\eta_2)^2 + (\overline{\lambda}_3 v^2)^2}$. The masses of the
charged components of $\Phi_{\pm}$ are $\overline{m}^2 - \lambda_2
v^2 \pm (\sigma_4 \eta_1 \eta_2 - \lambda_4 v^2)$. From this it can
be seen that splitting between charged and neutral components is
different for $\Phi_+$ and for $\Phi_-$.

The Higgs doublet $\Phi'$ is $CPS$-split. Altogether, then, there
are four distinct $SU(2)_L$-breaking splittings within the three
extra Higgs doublets. These are determined by the three parameters
$\lambda_i$, $i = 2,3,4$. There is therefore one relation.

\subsection{$G_{\Phi} = SU(N) \times SU(N)' \times S_2$, $R_{\Phi} =
(N,1) + (1,N)$.}

Here the group has two $SU(N)$ factors and a discrete symmetry under
which they are interchanged. We can write the messenger field as
$(\eta^{\alpha}, \eta^{\alpha'})$, in an obvious notation. Without
loss of generality, one can bring the VEV of the messenger field to
the form

\begin{equation}
\langle \eta^{\alpha} \rangle = \left( \begin{array}{c} \eta \\ 0 \\ : \\
0
\end{array} \right), \;\;\;\; \langle \eta^{\alpha'} \rangle =
\left( \begin{array}{c} \eta' \\ 0 \\ : \\ 0 \end{array} \right),
\end{equation}

\noindent where $\eta$ and $\eta'$ are real. The Yukawa terms are of
the form $Y_{mn} \psi_m \psi_n (\eta^*_{\alpha} \Phi^{\alpha} +
\eta^*_{\alpha'} \Phi^{\alpha'})$. So that $\Phi_F = \eta \Phi^1 +
\eta' \Phi^{1'}$.  The most general form of the Higgs-flavor
breaking part of the Higgs potential is

\begin{equation}
\begin{array}{cl}
V_{HFB} & = \sigma_1 \left( \Phi_{\alpha}^{\dag} \cdot \Phi^{\alpha}
\; \eta^*_{\beta} \eta^{\beta} + \Phi_{\alpha'}^{\dag} \cdot
\Phi^{\alpha'} \; \eta^*_{\beta'} \eta^{\beta'} \right) \\
& + \sigma_2 \left( \Phi_{\alpha}^{\dag} \cdot \Phi^{\alpha} \;
\eta^*_{\beta'} \eta^{\beta'} + \Phi_{\alpha'}^{\dag} \cdot
\Phi^{\alpha'} \; \eta^*_{\beta} \eta^{\beta} \right) \\
& + \sigma_3 \left( \Phi_{\alpha}^{\dag} \cdot \Phi^{\beta} \;
\eta^*_{\beta} \eta^{\alpha} + \Phi_{\alpha'}^{\dag} \cdot
\Phi^{\beta'} \; \eta^*_{\beta'} \eta^{\alpha'} \right) \\
& + \sigma_4 \left( \Phi_{\alpha}^{\dag} \cdot \Phi^{\beta'} \;
\eta^*_{\beta'} \eta^{\alpha} + \Phi_{\alpha'}^{\dag} \cdot
\Phi^{\beta} \; \eta^*_{\beta} \eta^{\alpha'} \right). \end{array}
\end{equation}

\noindent  Substituting the VEV given in Eq. (21) into Eq. (22), one
sees that the Higgs doublets $\Phi^{\alpha}$, $\alpha = 2, ...,N$
all have mass-squared $\sigma_1 \eta^2 + \sigma_2 \eta^{\prime 2}$
and the $\Phi^{\alpha'}$, $\alpha' = 2, ...,N$ all have mass-squared
$\sigma_1 \eta^{\prime 2} + \sigma_2 \eta^2$. This reflects the fact
that the form of the messenger VEV leaves a global $SU(N-1) \times
SU(N-1)'$ unbroken in the low-energy Higgs sector, but not the
discrete interchange symmetry $S_2$. These symmetries prevent any of
these $2(N-1)$ extra Higgs doublets from decaying entirely into
Standard Models particles. In fact, the lightest components of each
of these doublets is absolutely stable. The $\Phi^1$ and $\Phi^{1'}$
mix:

\begin{equation}
(\Phi^1, \Phi^{1'}) \left( \begin{array}{cc} M_0^2 + (\sigma_1 +
\sigma_3) \eta^2 + \sigma_2 \eta^{\prime 2} & \sigma_4 \eta \eta' \\
\sigma_4 \eta \eta' & M_0^2 + (\sigma_1 + \sigma_3) \eta^{\prime 2}
+ \sigma_2 \eta^2
\end{array}
\right) \left( \begin{array}{c} \Phi^1 \\ \Phi^{1'} \end{array}
\right).
\end{equation}

\noindent Thus, we may write $\Phi = \cos \theta \Phi^1 + \sin
\theta \Phi^{1'}$ and $\Phi' = -\sin \theta \Phi^1 + \sin \theta
\Phi^{1'}$, and both of these Higgs doublets couple to quarks and
leptons (and do so proportionally to each other). The lighter of
them (which one can take to be $\Phi$) must be identified with the
Standard Model Higgs doublet $\Phi_{SM}$. (Its mass-squared will be
the smallest of all the Higgs doublets if, say, $\sigma_3$ is
sufficiently negative.) The three distinct masses of the extra Higgs
doublets and the mixing angle $\theta$ are determined by the four
parameters, $\sigma_2 \eta^2$, $(\sigma_1 + \sigma_3)/\sigma_2$,
$\sigma_4/\sigma_2$, and $\eta'/\eta$; so that there is no relation
among them.

The $SU(2)_L$-breaking splittings within the Higgs doublets come
from

\begin{equation}
\begin{array}{cl}
V_4 & = \lambda_1 \left( \Phi_{\alpha}^{\dag} \cdot \Phi^{\alpha} \;
\Phi^{\dag}_{\beta} \cdot \Phi^{\beta} + \Phi_{\alpha'}^{\dag} \cdot
\Phi^{\alpha'} \; \Phi^{\dag}_{\beta'} \cdot \Phi^{\beta'} \right) \\
& + \lambda_2 \; \Phi_{\alpha}^{\dag} \cdot \Phi^{\alpha} \;
\Phi^{\dag}_{\beta'} \cdot \Phi^{\beta'} \\
& + \lambda_3 \left( \Phi_{\alpha}^{\dag} \cdot \Phi^{\beta} \;
\Phi^{\dag}_{\beta} \cdot \Phi^{\alpha} + \Phi_{\alpha'}^{\dag}
\cdot
\Phi^{\beta'} \; \Phi^{\dag}_{\beta'} \cdot \Phi^{\alpha'} \right) \\
& + \lambda_4 \; \Phi_{\alpha}^{\dag} \cdot \Phi^{\beta'} \;
\Phi^{\dag}_{\beta'} \cdot \Phi^{\alpha}.
\end{array}
\end{equation}

It is straightforward to show that $\Phi^{\alpha}$, $\alpha = 2,
..., N$ is $CN$-split and that the splitting is independent of
$\alpha$ (because of the residual symmetry). The same is true for
$\Phi^{\alpha'}$, $\alpha' = 2, ..., N$; but the splitting is not
the same for $\Phi^{\alpha}$ and $\Phi^{\alpha'}$. The sequential
Higgs doublet $\Phi'$ is $CPS$-split. Altogether, then, there are
four distinct $SU(2)_L$-breaking splittings, and they are determined
by the four parameters $\lambda_i$, $i = 1,..., 4$, so that there is
no relation among them.

The dark matter would consist of $2(N-1)$ complex neutral fields. Of
these, $N-1$ would be degenerate with one mass, and $N-1$ would be
degenerate with another mass.

\subsection{$G_{\Phi} = SO(N) \times SO(N)' \times S_2$, $R_{\Phi} =
(N,1) + (1,N)$.}

\subsubsection{$\eta$ = real.}

For $N \neq 3$ this case is very similar to the $SU(N) \times SU(N)'
\times S_2$ case considered above. The main difference concerns the
$SU(2)_L$-breaking splittings, which in this case depend on {\it
six} terms (as compared to the four in Eq. (24)):

\begin{equation}
\begin{array}{cl}
V_4 & = \lambda_1 \left( \Phi^{i \dag} \cdot \Phi^i \; \Phi^{j \dag}
\cdot \Phi^j + \Phi^{i' \dag} \cdot
\Phi^{i'} \; \Phi^{j' \dag} \cdot \Phi^{j'} \right) \\
& + \lambda_2 \; \Phi^{i \dag} \cdot \Phi^i \;
\Phi^{j' \dag} \cdot \Phi^{j'} \\
& + \lambda_3 \left( \Phi^{i \dag} \cdot \Phi^j \; \Phi^{j \dag}
\cdot \Phi^i + \Phi^{i' \dag} \cdot
\Phi^{j'} \; \Phi^{j' \dag} \cdot \Phi^{i'} \right) \\
& + \lambda_4 \left( \Phi^{i \dag} \cdot \Phi^j \; \Phi^{i \dag}
\cdot \Phi^j + \Phi^{i' \dag} \cdot \Phi^{j'} \; \Phi^{i' \dag}
\cdot \Phi^{j'} \right) \\
& + \lambda_5 \; \Phi^{i \dag} \cdot \Phi^{j'} \; \Phi^{j' \dag}
\cdot \Phi^i \\
& + \lambda_6 \left( \Phi^{i \dag} \cdot \Phi^{j'} \; \Phi^{i \dag}
\cdot \Phi^{j'} + \Phi^{i' \dag} \cdot \Phi^j \; \Phi^{i' \dag}
\cdot \Phi^j   \right).
\end{array}
\end{equation}

\noindent The $SU(2)_L$-breaking splittings produced by these terms
within the Higgs doublets $\Phi^i$, $i=2, .., N$ and $\Phi^{i'}$,
$i' = 2, ..., N$, now include the splitting of neutral scalars from
neutral pseudoscalars (caused by the $\lambda_4$ and $\lambda_6$
terms, which give mass to $(\Phi^i)^2 + h.c.)$, whereas the other
terms only give mass to $|\Phi^i|^2$). Thus, all the extra Higgs
doublets are $CPS$-split. Altogether, then, there are six distinct
$SU(2)_L$-breaking splittings (two within $\Phi'$; two within
$\Phi^i$, $i=2,...,N$; and two within $\Phi^{i'}$, $i' = 2, ...,N$).
Since these depend on the six parameters $\lambda_i$, $i = 1,...,6$,
there are no relations among these splittings.

The dark matter would consist of $2(N-1)$ real neutral fields. Of
these, $N-1$ would be degenerate with one mass, and $N-1$ would be
degenerate with another mass.

The case $N=3$ is special, as it was for $SO(N)$ (section 3.2.2),
because there is an extra term in $V_{HFB}$ due to the rank-3
epsilon symbol:

\begin{equation}
\begin{array}{cl}
V_{HFB} & = \sigma_1 \left( \Phi^{i \dag} \cdot \Phi^i \; \eta^j
\eta^j + \Phi^{i' \dag} \cdot
\Phi^{i'} \; \eta^{j'} \eta^{j'} \right) \\
& + \sigma_2 \left( \Phi^{i \dag} \cdot \Phi^i \; \eta^{j'}
\eta^{j'} + \Phi^{i' \dag} \cdot
\Phi^{i'} \; \eta^j \eta^j \right) \\
& + \sigma_3 \left( \Phi^{i \dag} \cdot \Phi^j \; \eta^j \eta^i +
\Phi^{i' \dag} \cdot
\Phi^{j'} \; \eta^{j'} \eta^{i'} \right) \\
& + \sigma_4 \left( \Phi^{i \dag} \cdot \Phi^{j'} \; \eta^{j'}
\eta^i + \Phi^{i' \dag} \cdot \Phi^j \; \eta^j \eta^{i'} \right) \\
& + i m_5 \left( \Phi^{i \dag} \cdot \Phi^j \eta^k \epsilon^{ijk} +
\Phi^{i' \dag} \cdot \Phi^{j'} \eta^{k'} \epsilon^{i' j' k'}
\right).
\end{array}
\end{equation}

\noindent The last term has the effect of mixing $\Phi^2$ with
$\Phi^3$ and also $\Phi^{2'}$ with $\Phi^{3'}$. The mass-squared
eigenstates are as follows: (1) $\Phi^1$ and $\Phi^{1'}$ mix,
similarly to Eq. (23). One linear combination of them is $\Phi_{SM}$
and the other is a sequential Higgs doublet $\Phi'$. (2) $\Phi^2$
and $\Phi^3$ mix, giving eigenstates $\Phi_{\pm} =
\frac{1}{\sqrt{2}} (\Phi^2 \pm i \Phi^3)$. (3) $\Phi^{2'}$ and
$\Phi^{3'}$ mix, giving eigenstates $\Phi'_{\pm} =
\frac{1}{\sqrt{2}} (\Phi^{2'} \pm i \Phi^{3'})$. The five masses of
the extra Higgs doublets and the mixing angle of $\Phi'$ are
determined by the five coefficients in $V_{HFB}$ and the ratio of
the VEVs of $\eta^1$ and $\eta^{1'}$. Thus, there is no relation
among them.

The fields $\Phi_{\pm}$ and $\Phi'_{\pm}$ are $CN$-split, as was the
case for $SO(3)$ in section 3.2.2. $\Phi'$ is $CPS$-split. The six
distinct splittings are determined by the six coefficients in $V_4$,
shown in Eq. (25). So, again, there are no relations among them.

\subsubsection{$\eta$ = complex.}

For $\eta$ complex (and charged under some $U(1)$), $N=3$ is not a
special case, as there is no extra cubic term involving the epsilon
symbol. (It is forbidden by the $U(1)$.)

The messenger field VEV for complex $\eta$ cannot in general be
brought to the form given in Eq. (21), but can be brought to the
form

\begin{equation}
\langle \eta^i \rangle = \left( \begin{array}{c} \eta_1 \\ i \eta_2 \\ 0 \\ : \\
0
\end{array} \right), \;\;\;\; \langle \eta^{i'} \rangle =
\left( \begin{array}{c} \eta'_1 \\ i \eta'_2 \\ 0 \\ : \\ 0
\end{array} \right),
\end{equation}

\noindent One consequence of this is that $\Phi_F$ is a linear
combination of $\Phi^1$, $\Phi^2$, $\Phi^{1'}$, and $\Phi^{2'}$.
$V_{HFB}$ mixes these to give four doublets, $\Phi_{SM}$, and three
sequential Higgs doublets $\Phi'$, $\Phi^{\prime \prime}$, and
$\Phi^{\prime \prime \prime}$). The VEV in Eq. (27) leaves unbroken
an $SO(N-2) \times SO(N-2)$. Because of this, the $\Phi^i$, $i = 3,
..., N$ are degenerate, as are the $\Phi^{i'}$, $i' = 3, ..., N$
(though they are not degenerate with each other). There are thus
five distinct masses for the extra Higgs doublets, and three mixing
angles (of $\Phi'$, $\Phi^{\prime \prime}$, and $\Phi^{\prime \prime
\prime}$ with $\Phi_F$). These are determined by eight parameters:
five couplings $\sigma_i$, $i = 2,...,6$, in $V_{HFB}$, which has a
structure analogous to that shown in Eq. (25), and three ratios of
the VEVs given in Eq. (27). So, again, there is no relation among
these quantities.

On the other hand, there are now 10 distinct $SU(2)_L$-breaking
splittings {\it within} the extra Higgs doublets, since all five
types of extra Higgs doublet ($\Phi'$, $\Phi^{\prime \prime}$,
$\Phi^{\prime \prime \prime}$, $\Phi^i$, and $\Phi^{i'}$) are
$CPS$-split. These ten quantities are determined by the six
parameters $\lambda_i$, $i = 1,...,6$ shown in Eq. (25). Thus, there
are four relations.

The dark matter would consist of $2(N-2)$ real neutral fields. Of
these, $N-2$ would be degenerate with one mass, and $N-2$ would be
degenerate with another mass.

\subsection{$G_{\Phi} = SU(2)$, $R_{\Phi} = N$}

The cases $N=2$ and $N=3$ have already been covered: the former in
subsection 3.1 with $N=2$, and the latter in subsections 3.2.2 and
3.3.1 with $N=3$.

\subsubsection{$N=4$ or $6$.}

In the case $N=4$ case, both $\Phi$ and $\eta$ are in a ${\bf 4}$ of
$SU(2)$. The VEV of $\eta$ cannot be brought to a very simple form
by a choice of $SU(2)$ basis; rather at least three of its
components remain non-zero. It turns out, therefore, that all four
Higgs doublets mix, in the sense that $\Phi_F$ is a linear
combination of all four mass-squared eigenstates of $V_{HFB}$. Thus
they are all sequential Higgs doublets that are unstable to decay
into quarks and leptons. None contain stable components that could
be dark matter. The number of masses of extra Higgs doublets is
three and the number of mixing angles for the extra Higgs is three.
These 6 quantities are determined by 7 parameters (3 coefficients in
$V_{HFB}$ and 4 real parameters from the form of the messenger VEV),
so that there are no relations among them. There are testable
relations, however, if one also takes into account the
$SU(2)_L$-breaking splittings within the Higgs doublets. There are 6
such splittings (two within each extra Higgs doublet, and three
extra Higgs doublets). These are produced by $V_4$, which has four
terms, since in $\Phi^{\dag} \cdot \Phi \eta^{\dag} \eta$ the
product $\eta^{\dag} \eta$ must be in ${\bf 4} \times {\bf 4} = {\bf
1} + {\bf 3} + {\bf 5} + {\bf 7}$. Of these four terms, three
contribute to splittings. Altogether, then, one has 12 quantities
determined by 10 parameters, giving two relations.

The $N=6$ case is much like the $N=4$ case. Again, the form of the
messenger VEV cannot be made very simple, and all the extra Higgs
doublets mix with $\Phi_F$ and are thus sequential Higgs doublets
that can decay into quarks and leptons. Again, it turns out that
taking into account the $SU(2)_L$-breaking splittings within the
five extra Higgs doublets, there are two relations.

\subsubsection{$N=5$, $\eta$ = real.}

In this case, both $\Phi$ and $\eta$ are in the ${\bf 5}$ of $SU(2)$
(or equivalently $SO(3)$). Since this case was studied in detail in
\cite{Barr:2011yq}, here we will only summarize the results. The
${\bf 5}$ is the rank-2 symmetric traceless representation of
$SO(3)$. Thus we may write the Higgs and messenger fields as
$\Phi^{(ij)}$ and $\eta^{(ij)}$, where $i,j = 1,2,3$. Since the VEV
of the messenger field is a real symmetric traceless matrix, it can
be diagonalized by a choice of $SO(3)$ basis:

\begin{equation}
\langle \eta^{(ij)} \rangle = \left( \begin{array}{ccc} a +
\frac{1}{\sqrt{3}} b & 0 & 0 \\ 0 & - a + \frac{1}{\sqrt{3}} b & 0
\\ 0 & 0 & - \frac{2}{\sqrt{3}} b \end{array} \right).
\end{equation}

The Higgs fields $\Phi^{(ij)}$ can also be thought of as a symmetric
traceless matrix. In the basis where the messenger VEV has the form
in Eq. (28), one can distinguish the three ``off-diagonal" Higgs
doublets, $\Phi^{(ij)}$, with $i \neq j$, from the two ``diagonal"
Higgs doublets. The mass-squared matrix produced by $V_{HFB}$ leads
to the result that the mass-squared eigenstates are $\Phi^{(12)}$,
$\Phi^{(23)}$, and $\Phi^{(31)}$ and two linear combinations of the
diagonal Higgs doublets, which one may call $\Phi = \Phi_{SM}$ and
$\Phi'$. One finds that $\Phi_F = \cos \theta \Phi_{SM} + \sin
\theta \Phi'$, where in general $\theta$ is some non-trivial angle.
Thus, in general, $\Phi'$ is a sequential Higgs doublet, and is
unstable.

In $V_{HFB}$ there are three independent terms of the form
$\Phi^{\dag} \cdot \Phi \eta \eta$, since $\eta \eta$ must be in the
symmetric product $[{\bf 5} \times {\bf 5}]_S = {\bf 1} + {\bf 5} +
{\bf 9}$. (Only the two non-singlet contractions contribute to
splitting within the ${\bf 5}$ of Higgs doublets, however.) There
can also be a cubic term of the form $\Phi^{\dag} \cdot \Phi \eta$,
since $\Phi^{\dag} \cdot \Phi$ is in $[{\bf 5} \times {\bf 5}]$,
which contains a ${\bf 5}$. Thus, the masses of the four extra Higgs
doublets and the mixing angle $\theta$ depend on only 4 parameters
(three coefficients within $V_{HFB}$ and the ratio $b/a$ in the
messenger VEV). Thus, there is one prediction. For example, if one
knew the masses of the four extra Higgs doublets, one could predict
the angle $\theta$, and thus know the strength of the Yukawa
couplings of $\Phi'$ to the quarks and leptons.

The form of the messenger VEV leaves unbroken discrete symmetries
that stabilize some of the Higgs fields. Defining $P_{23}$ as a
rotation by $\pi$ in the 23 plane (of $SO(3)$), which is equivalent
to reflections in the 2 and 3 directions, one has that $\Phi^{(23)}$
is even under it, while $\Phi^{(12)}$ and $\Phi^{(31)}$ are odd.
Similarly, one can define $P_{12}$ and $P_{31}$. These symmetries
prevent any of the off-diagonal Higgs doublets from decaying into
just Standard Models fields (which are all even under them). They
do, however, allow the heaviest of the three off-diagonal Higgs
doublets to decay into the other two plus Standard Model fields (via
terms like $\Phi^{(12) \dag} \cdot \Phi^{(23)} \Phi^{(31) \dag}
\cdot \Phi^{(11)}$), if that is kinematically allowed. Thus, the two
lighter off-diagonal Higgs doublets have absolutely stable
components. For certain values of the parameters, all three of the
off-diagonal Higgs doublets have stable components.

An interesting special case arises if the cubic term in $V_{HFB}$
vanishes, as may happen due to a symmetry under which the messenger
field is odd. In this case, the angle $\theta$ is zero, and the
diagonal extra Higgs doublet $\Phi'$ does not couple to quarks and
leptons, and has stable components.

The extra Higgs doublets are all $CN$-split. Since there are four
extra Higgs doublets, there are altogether eight distinct
$SU(2)_L$-breaking splittings. These depend on just four
coefficients in $V_4$. (In $V_4$, there are five independent terms,
since $\Phi^{\dag} \cdot \Phi$ must be in ${\bf 5} \times {\bf 5} =
{\bf 1} + {\bf 3} + {\bf 5} + {\bf 7} + {\bf 9}$. The four terms
corresponding to non-singlet contractions contribute to the
splittings.) Thus there are four relations among the
$SU(2)_L$-breaking splittings.

The dark matter consists of two, three, or four real fields of
different masses, depending on the values of model parameters.

\subsubsection{$N=5$, $\eta$ = complex.}

Here we assume that there is a $U(1)$ under which $\Phi$ and $\eta$
have the same non-zero charge. The big difference with the previous
case (of $\eta$ real), is that the messenger VEV, being a complex
symmetric matrix, can no longer be diagonalized by $SO(3)$
transformations. As a result, $\Phi_F$ is a linear combination of
all five of the Higgs doublets. Consequently, they are all unstable
to decay into quarks and leptons. The four mass-squared eigenvalues
of the extra Higgs doublets, and their four mixing angles with
$\Phi_F$, altogether 8 quantities, are determined by 9 parameters,
so that there are no relations among them. (The 9 parameters are 4
coefficients in $V_{HFB}$ and 5 from the form of $\langle \eta
\rangle$.)

When the 8 $SU(2)_L$-breaking splittings are included, and the 4
parameters in $V_4$, one has altogether 16 quantities determined by
13 parameters. So, overall, there are 3 relations among masses and
mixings.

\subsubsection{$G_{\Phi} = SO(5)$,  $R_{\Phi} = 4$.}

The group $SO(5)$ has a four-component spinor representation. One
can show that if the messenger field is in the spinor, then its VEV
can be brought to a form that breaks $SO(5)$ down to $SU(2)$, where
$SU(2) \subset SU(2) \times SU(2)' \equiv SO(4) \subset SO(5)$.
(This breaks seven generators of $SO(5)$, and the rotations
associated with these seven generators allow one to eliminate seven
of the eight real parameters in $\langle \eta \rangle$.) Under the
unbroken $SU(2)$ subgroup, the ${\bf 4}$ Higgs doublets decompose
into ${\bf 1} + {\bf 1} + {\bf 2} = \Phi^{SM} + \Phi' +
\Phi^{\alpha}$, $\alpha = 1,2$. One finds that $\Phi_F$ is a mixture
of $\Phi'$ and $\Phi_{SM}$, so that $\Phi'$ is a sequential Higgs
doublet. $\Phi^{\alpha}$ is a degenerate pair of inert Higgs
doublets. The counting of parameters is easy to do, using the
methods used above and the fact that ${\bf 4} \times {\bf 4} = {\bf
1} + {\bf 5} + {\bf 9}$. One finds that there are no predictions for
the masses or mixing angle.

\subsection{Other Cases}

Here we deal with other cases that involve continuous groups, all of
which accommodate six Higgs doublets.

\subsubsection{$G_{\Phi} = SU(3)$, $R_{\Phi} = 6$.}

This case is quite similar to that considered in section 3.6.2
($SO(3)$ with $R_{\Phi} = 5$). The ${\bf 6}$ of $SU(3)$ is a
symmetric rank-2 (traceful) tensor, so that we may write the Higgs
and messenger fields as $\Phi^{(\alpha \beta)}$ and $\eta^{(\alpha
\beta)}$. Without loss of generality, the messenger VEV can be
brought to the form

\begin{equation}
\langle \eta^{(\alpha \beta)} \rangle = \left( \begin{array}{ccc} a
& 0 & 0 \\ 0 & b & 0
\\ 0 & 0 & c \end{array} \right),
\end{equation}

\noindent where $a$, $b$, and $c$ are real. $\Phi_F \propto a
\Phi^{(11)} + b \Phi^{(22)} + c \Phi^{(33)}$. Since $\overline{{\bf
6}} \times {\bf 6} = {\bf 1} + {\bf 8} + {\bf 27}$, there are three
independent terms in $V_{HFB}$:

\begin{equation}
V_{HFB} = \sigma_1 \Phi^{\dag}_{(\alpha \beta)} \cdot \Phi^{(\alpha
\beta)} \eta^*_{(\gamma \delta)} \eta^{(\gamma \delta)} + \sigma_2
\Phi^{\dag}_{(\alpha \beta)} \cdot \Phi^{(\beta \gamma)}
\eta^*_{(\gamma \delta)} \eta^{(\delta \alpha)} + \sigma_3
\Phi^{\dag}_{(\alpha \beta)} \cdot \Phi^{(\gamma \delta)}
\eta^*_{(\gamma \delta)} \eta^{(\alpha \beta)}.
\end{equation}

As in the case discussed in section 3.6.2, one can distinguish the
``off-diagonal" and ``diagonal" Higgs doublets. The off-diagonal
Higgs doublets $\Phi^{(12)}$, $\Phi^{(23)}$, and $\Phi^{(31)}$ are
eigenstates of the mass-squared matrix produced by $V_{HFB}$ with
eigenvalues $m^2 + \sigma_2 (a^2 + b^2)$, $m^2 + \sigma_2 (b^2 +
c^2)$, and $m^2 + \sigma_2 (c^2 + a^2)$, where $m^2 \equiv M_0^2 +
\sigma_1 (a^2 + b^2 + c^2)$.  The three diagonal Higgs doublets mix
with each other, i.e. $\Phi_F$ is a linear combination of all of
them, so that all of them couple to quarks and leptons and are
unstable. The five masses of the extra Higgs doublets and the two
mixing angles (of the diagonal Higgs doublets with $\Phi_F$), are
determined by four parameters ($\sigma_2 a^2$, $\sigma_3/\sigma_2$,
$b/a$, and $c/a$). There are therefore three relations among them.

As in section 3.6.2, there are discrete symmetries (including the
$P_{23}$, $P_{12}$, $P_{31}$, defined there) left unbroken by the
messenger VEV. These prevent any of the off-diagonal Higgs doublets
from decaying entirely to Standard Model fields. However, the
heaviest of the off-diagonal Higgs doublets can decay into the
lighter ones plus other particles, if this kinematically allowed.
Thus, at least two, and possibly all three of the off-diagonal Higgs
doublets have absolutely stable components, depending on the values
of parameters.

The potential $V_4$, which is responsible for the $SU(2)_L$-breaking
splittings within the extra Higgs doublets, contains three terms
(because the decomposition $\overline{{\bf 6}} \times {\bf 6} = {\bf
1} + {\bf 8} + {\bf 27}$ contains three terms), of which two
contribute to splittings. All of the extra Higgs doublets are
$CPS$-split. (For example, the scalar and pseudoscalar components of
$\Phi^{(12)}$ are split from each other by such terms as $\langle
\Phi^{\dag}_{(11)} \rangle \cdot \Phi^{(12)} \langle
\Phi^{\dag}_{(22)} \rangle \cdot \Phi^{(12)} + h.c.$) Altogether,
there are ten $SU(2)_L$-breaking splittings within the five extra
Higgs doublets, and these are determined by coefficients in $V_4$,
so that there are eight relations among these splittings.

The dark matter consists of at least two, and possibly three, real
neutral fields that have different masses.

\subsubsection{$G_{\Phi} = SU(2) \times SU(2)' \times SU(2)^{\prime \prime} \times
S_3$, $R_{\Phi} = (2,1,1) + (1,2,1) + (1,1,2)$.}

In this case there are three $SU(2)$ groups and a permutation
symmetry that interchanges them. In an obvious notation, the
messenger field can be written $(\eta^{\alpha}, \eta^{\alpha'},
\eta^{\alpha^{\prime \prime}})$, with the indices taking the values
1,2. The messenger VEV can be brought to the form in which

\begin{equation}
\langle \eta^{\alpha} \rangle = \left( \begin{array}{c} a \\ 0
\end{array} \right), \;\;   \langle \eta^{\alpha'} \rangle =
\left( \begin{array}{c} b \\ 0 \end{array} \right), \;\; \langle
\eta^{\alpha^{\prime \prime}} \rangle = \left( \begin{array}{c} c \\
0
\end{array} \right),
\end{equation}

\noindent This gives $\Phi_F \propto a \Phi^1 + b \Phi^{1'} + c
\Phi^{1^{\prime \prime}}$.

The potential $V_{HFB}$ contains three terms: $\sigma_1
(\Phi^{\dag}_{\alpha} \cdot \Phi^{\alpha} \eta^*_{\beta}
\eta^{\beta} + perm.) + \sigma_2 (\Phi^{\dag}_{\alpha} \cdot
\Phi^{\alpha} \eta^*_{\beta'} \eta^{\beta'} + perm.) + \sigma_3
(\Phi^{\dag}_{\alpha} \cdot \Phi^{\beta'} \eta^*_{\beta'}
\eta^{\alpha} + perm.)$. This produces a mass-squared matrix, under
which the three Higgs doublets $\Phi^2$, $\Phi^{2'}$, and
$\Phi^{2^{\prime \prime}}$ are eigenstates. The other eigenstates
(which we may call $\Phi = \Phi_{SM}$, $\Phi'$, and $\Phi^{\prime
\prime}$) are linear combinations of $\Phi^1$, $\Phi^{1'}$, and
$\Phi^{1^{\prime \prime}}$, which all mix, in general, with $\Phi_F$
and therefore couple to quarks and leptons and are unstable. There
are five mass-squareds of the extra Higgs doublets, and two mixing
angles (of the extra sequential Higgs with $\Phi_F$), which depend
on 5 parameters ($\sigma_1 a^2$, $\sigma_2/\sigma_1$,
$\sigma_3/\sigma_1$, $b/a$, $c/a$), giving two relations.

The messenger VEV leaves unbroken discrete symmetries $P_2$,
$P_{2'}$, $P_{2^{\prime \prime}}$ (where $P_2$ is a reflection in
the 2 direction in $SU(2)$, etc.), which make the lightest
components of each of the Higgs doublets $\Phi^2$, $\Phi^{2'}$, and
$\Phi^{2^{\prime \prime}}$ absolutely stable.

There are three terms in $V_4$ (whose form is analogous to that of
$V_{HFB}$). It can be shown that these cause $\Phi'$ and
$\Phi^{\prime \prime}$ to be $CPS$-split, but $\Phi^2$, $\Phi^{2'}$,
and $\Phi^{2^{\prime \prime}}$ to be $CN$-split. There are therefore
a total of 7 distinct $SU(2)_L$-breaking splittings, which are
determined by 3 parameters (the coefficients in $V_4$), to give 4
relations.

The dark matter particles would consist of three charged neutral
fields of different masses, namely the neutral components of
$\Phi^2$, $\Phi^{2'}$, and $\Phi^{2^{\prime \prime}}$.

\subsubsection{$G_{\Phi} = SU(3) \times SU(2)$, $R_{\Phi} = (3,2)$.}

Denoting the $SU(3)$ indices by $\alpha$, $\beta$, etc., and the
$SU(2)$ indices by $\lambda$, $\mu$, etc., one can write the Higgs
and messenger fields as $\Phi^{\alpha \lambda}$, $\eta^{\alpha
\lambda}$. Without loss of generality, one can bring the messenger
field VEV to the form

\begin{equation}
\langle \eta^{\alpha \lambda} \rangle = \left( \begin{array}{cc}
\eta & 0 \\ 0 & \kappa \\ 0 & 0 \end{array} \right),
\end{equation}

\noindent where $\eta$ and $\kappa$ are real. Then $\Phi_F = \eta
\Phi^{11} + \kappa \Phi^{22}$. Since in $\Phi^{\dag} \cdot \Phi
\eta^{\dag} \eta$, the product $\eta^{\dag} \eta$ must be in
$(3,2)^* \times (3,2) = (1,1) + (1,3) + (8,1) + (8,3)$, there are
four independent terms in $V_{HFB}$:

\begin{equation}
\begin{array}{cl}
V_{HFB} & = \sigma_1 \Phi^{\dag}_{\alpha \lambda} \cdot \Phi^{\alpha
\lambda} \eta^*_{\beta \mu} \eta^{\beta \mu} \\
& + \sigma_2 \Phi^{\dag}_{\alpha \lambda} \cdot \Phi^{\beta
\lambda} \eta^*_{\beta \mu} \eta^{\alpha \mu} \\
& + \sigma_3 \Phi^{\dag}_{\alpha \lambda} \cdot \Phi^{\alpha
\mu} \eta^*_{\beta \mu} \eta^{\beta \lambda} \\
& + \sigma_4 \Phi^{\dag}_{\alpha \lambda} \cdot \Phi^{\beta \mu}
\eta^*_{\beta \mu} \eta^{\alpha \lambda}.
\end{array}
\end{equation}

The mass-squared matrix produced by this has the following
eigenstates: $\Phi = \Phi_{SM} = \cos \theta \Phi^{11} + \sin \theta
\Phi^{22}$, $\Phi' = -\sin \theta \Phi^{11} + \cos \theta
\Phi^{22}$, $\Phi^{12}$, $\Phi^{21}$, $\Phi^{31}$, and $\Phi^{32}$.
The five masses of the extra Higgs doublets and the angle $\theta$
are determined by four parameters ($\sigma_i \eta^2$, $i=1,..,4$,
and $\kappa/\eta$), so that there are two relations.

The fields $\Phi^{31}$ and $\Phi^{32}$ are $CN$-split, whereas the
other extra Higgs doublets are all $CPS$-split. Thus there are 8
distinct $SU(2)_L$-breaking splittings, which are determined by 3
coefficients in $V_4$, giving 5 relations among them.

Left unbroken by the messenger VEV is a discrete symmetry $P_3$ that
is a reflection in 3 direction of $SU(3)$. Under this symmetry,
$\Phi^{31}$ or $\Phi^{32}$ are odd, while all other fields are even.
So neither $\Phi^{31}$ nor $\Phi^{32}$ can decay entirely into
Standard Model particles, and the lighter of the two contains an
absolutely stable component (a complex neutral field). The quartic
terms $\Phi^{\dag}_{32} \cdot \Phi^{31} \Phi^{\dag}_{21} \cdot
\Phi^{22}$ and $\Phi^{\dag}_{32} \cdot \Phi^{31} \Phi^{\dag}_{11}
\cdot \Phi^{12}$ allow the heavier of $\Phi^{31}$ and $\Phi^{32}$ to
decay to the lighter of them plus $\Phi^{21}$ or $\Phi^{12}$ or
$\Phi^*_{21}$ or $\Phi^*_{12}$ plus Standard Model fermions, if this
is kinematically possible. If none of these decays are kinematically
possible (because $\Phi^{12}$ and $\Phi^{21}$ are too heavy) then
{\it both} $\Phi^{31}$ and $\Phi^{32}$ contain absolutely stable
complex neutral fields.

The quartic term $\Phi^{\dag}_{11} \cdot \Phi^{12} \Phi^{\dag}_{22}
\cdot \Phi^{21}$ allows the heavier of $\Phi^{12}$ and $\Phi^{21}$
to decay to the lighter one plus Standard Model fermions. If the
lighter one is sufficiently heavy, then it can in turn decay into
$\Phi^{31} + \Phi^*_{32}$ + fermions or into $\Phi^{32} +
\Phi^*_{31}$ + fermions, via the quartic terms mentioned in the
previous paragraph. Otherwise, the lighter of $\Phi^{12}$ and
$\Phi^{21}$ contains an absolutely stable field, namely a real
neutral field.

In short, depending on the values of the masses, there can be the
following absolutely stable particles: (i) two complex neutral
fields and a real neutral field, (ii) two complex neutral fields, or
(iii) one complex neutral field and a real neutral field.

\subsubsection{$G_{\Phi} = SU(2) \times SU(2)$, $R_{\Phi} = (3,2)$.}

In this case, the VEV of the messenger field cannot be brought to a
simple form, so that it turns out that all the extra Higgs fields
mix with $\Phi_F$, and all are unstable to decay to Standard Model
fermions. However, there are some relations among the masses and
mixings.

\section{Discrete Higgs-flavor Groups}

So far we have only discussed cases where the non-abelian
Higgs-flavor group contains continuous symmetries. Another logical
possibility is that the Higgs-flavor group has no continuous
factors, but is a discrete group $D_{\Phi}$. (For a discussion of
non-abelian discrete symmetries in particle physics, see
\cite{Ishimori:2010au}.) In the Appendix, we briefly discuss some of
the group theory for all the cases where the order of $D_{\Phi}$ is
less than 16 and a few other cases. One finds from this analysis
that with $o(D_{\Phi}) < 16$ there are relatively few
distinguishable cases. The largest $D_{\Phi}$ multiplets are
3-dimensional. And many of the groups give similar structure to the
Higgs potential. Aside from this, discrete Higgs-flavor groups are
less interesting than continuous ones for two reasons.

First, if the Higgs-flavor group is discrete, the messenger field
VEV cannot in general be brought to a simple form by $D_{\Phi}$
transformations. This in practice means that $\Phi_F$ ends up being
a linear combination of all the Higgs doublets. Therefore, all the
extra Higgs doublets are unstable to decay to quarks and leptons,
and none can play the role of dark matter.

Second, by imposing a discrete symmetry we eliminate various terms
from the Higgs potential $V_H$ and impose relations on the coupling
constants of many of the others. These constraints often result in
additional accidental continuous symmetries of $V_H$. However, since
$D_{\Phi}$ and its associated accidental symmetry $\Delta_{\Phi}$ is
not gauged, the breaking of $\Delta_{\Phi}$ results in
pseudo-Goldstone bosons in the spectrum. Moreover, if discrete
symmetries are not protected by being gauged, they are subject to
being violated by gravity. The Peccei-Quinn solution to the strong
CP problem is such an example where gravity spoils the solution
\cite{Holman:1992us}.

\section{Conclusions}

We have shown that there are many simple possibilities for the
Higgs-flavor symmetry and representations. Moreover, the various
possibilities give quite distinctive spectra and properties for the
extra Higgs doublets. Indeed, it seems that hardly any two cases are
exactly alike. They differ in the number of extra Higgs doublets
there are; how many are ``sequential" and how many are ``inert";
whether the neutral components of the inert Higgs doublets are split
(``CPS splitting") or not (``CN" splitting); how many of the inert
Higgs fields are stable against decay to other Higgs fields; and the
group-theoretical relations among the masses of the extra Higgs
fields. the phenomenological possibilities are clearly rich, not
only for collider physics, but also for dark matter.

\section*{Appendix on discrete Higgs-flavor symmetry}

For models with discrete Higgs symmetry $D_{\Phi}$, it is difficult
to work out all the cases of extra Higgs doublets in $N$ dimensional
irreps of $D_{\Phi}$, even when $N$ is small. There are an infinite
number of such models, so that in order to make the analysis
manageable one needs to restrict the order of $D_{\Phi}$ or else
restrict to cases where the group $D_{\Phi}$ is in a series like the
dihedral groups $D_n$ or the dicyclic groups $Q_n$. These series all
have two dimensional irreps for arbitrary $n$, and would lead to
Higgs potentials that could easily be classified, but which will not
be classified here.

We will therefore consider only cases with $o(D_{\Phi}) \leq$ 16.
With $o(D_{\Phi}) < 16$ the maximum dimension on any irrep is 3.
Abelian groups need not be considered, since they only have 1-dim
irreps. [It should be noted, however, that 1-dim irreps of
non-abelian discrete symmetries may play a role if the discrete
group has to be ``gauged" and anomaly free \cite{Luhn:2008sa}). For
example, to embed an $S_3$ model in $SO(3)$ would require a singlet
since $3 \rightarrow 1' + 2$ when $SO(3) \rightarrow S_3$, where the
irreps of $S_3$ are $1,1'$ and 2. Likewise, to embed, say,
$\Delta_{27}$ in $SU(3)$ then singlets can arise in the
decomposition of the $SU(3)$ irreps, so that singlets are required
to construct anomaly free irreps of $\Delta_{27}$.]

It is not too difficult to work out the group products and construct
the Higgs potentials for all the non-abelian cases with $o(D_{\Phi})
< $ 16. We will not analyze the models, but make some
group-theoretical remarks relevant to classifying the models.

\vspace{0.2cm}

\noindent $o(G_{\Phi})=6$

\vspace{0.2cm} $S_3$ is the only non-abelian group of order 6. This
group contains a the irreps $1, 1', 2$. The product $2 \times 2$ has
the decomposition $2 \times 2= 1+1'+2$. There are consequently two
ways to (symmetrically) contract $\eta^2$ in the quartic terms of
form $\Phi^{\dag} \cdot \Phi \eta^2$ in $V_{HFB}$. Since $2 \times 2
\times 2$ contains a singlet, there is also a cubic term of the form
$\Phi^{\dag} \cdot \Phi \eta$ in $V_{HFB}$.

There is only one group or order 7 and it is abelian (in fact, all
groups of prime order are abelian), so the next cases to consider
are at order 8.

\vspace{0.2cm}

\noindent $o(G_{\Phi})=8$

\vspace{0.2cm}

There are two non-abelian groups to consider at order 8.

\vspace{0.2cm}

\noindent $D_4$: The order 8 dihedral group $D_4$ has four 1-dim
irreps ($1,1',1'',1'''$) and a single 2-dim irrep, with $(2 \times
2)_A = 1'$ and $(2 \times 2)_S = 1 + 1'' + 1'''$. This means that
there are three symmetric ways to contract $\eta^2$ and thus three
terms of the form $\Phi^{\dag} \cdot \Phi \eta^2$ in $V_{HFB}$.
Since $2\times 2\times 2$  contains no singlets, there is no cubic
term in $V_{HFB}$. Similar results hold for the $\Phi^4$ quartic
terms.

\vspace{0.2cm}

\noindent $Q_4$: The other group at order 8 is the first of the
dicyclic groups  $Q_4$  (also called the group of unit quaternions).
Like $D_4$, $Q_4$ has four 1-dim irreps ($1,1',1'',1'''$) and a
single 2-dim irrep, but they have different multiplication rules.
However, the only difference is in the products of 1-dim irreps, so
since we only consider 2-dim irreps for the $\Phi$ and $\eta$
fields, one gets exactly the same model as for $D_4$.

There are only abelian groups at order 9, so the next cases are at
order 10.

\vspace{0.2cm}

\noindent $o(G_{\Phi}) = 10$

\vspace{0.2cm}

\noindent $D_5$: This is the only non-abelian group at this order.
It has irreps $1,1',2,2'$. ($D_5$ will give models very much like
$S_3$, which is not surprising since $S_3$ is also a dihedral group,
i.e., $S_3=D_3$.)

If the model has only the 2, then since $(2\times 2)_S = 1 + 2$, and
$(2 \times 2)_A = 1'$, there will be two ways to symmetrically
contract $\eta^2$ and thus two terms of the form $\Phi^{\dag} \cdot
\Phi \eta^2$ in $V_{HFB}$. And since $(2 \times 2 \times 2)_S =
1+...$ there will be a single cubic term in $V_{HFB}$.

If the model has $2'$ instead of the 2, then due to the fact that
$(2' \times 2')_S = 1 + 2'$, one obtains the same model.

Since 11 is prime, there are only abelian groups at that order, and
the next cases are at order 12.

\vspace{0.2cm}

\noindent$o(G_{\Phi}) = 12$

\vspace{0.2cm}

There are three nonabelian groups to consider at this order, $T$,
$D_6$, and $Q_6$.

\vspace{0.2cm}  \noindent $T$: The tetrahedral group $T$ has irreps
$1,1',1'',3$, so the Higgs fields and the messenger field must be in
3. As $(3 \times
 3)_S = 1 + 1'+ 1''+3$ and $(3 \times 3)_A = 3$, $\eta^2$ can be
symmetrically contracted in four ways and there will be four terms
of the form $\Phi^{\dag} \cdot \Phi \eta^2$ in $V_{HFB}$, as well as
a single cubic term.

\vspace{0.2cm}

\noindent $D_6$: The irreps are $1_1, 1_2, 1_3, 1_4, 2, 2'$. There
are two possible models.

(i) With $\Phi$ and $\eta$ in the 2, one has, since $(2 \times 2)_S
= 1_1 + 2'$, two ways to symmetrically contract $\eta^2$ and two
terms of the form $\Phi^{\dag} \cdot \Phi \eta^2$ in $V_{HFB}$.
Since $2 \times 2'= 1_2 + 1_3 + 2$ there is no cubic term in
$V_{HFB}$.

(ii) With $\Phi$ and $\eta$ in the $2'$, one has $(2'\times 2')_S=
1_1 + 2'$, so that again there are two ways to symmetrically
contract $\eta^2$ and two terms of the form $\Phi^{\dag} \cdot \Phi
\eta^2$ in $V_{HFB}$. But now, since $2 \times 2'= 1_2 + 1_3 + 2$,
there is a cubic term.

\vspace{0.2cm}

\noindent $Q_6$: The irreps are $1_1,1_2,1_3,1_4,2,2'$, but with
slightly different multiplication rules. Since, however, the doublet
sector multiplications are the same as for $D_6$, one gets two
models identical to the (i) and (ii) models of $D_6$.

Since 13 is prime, the next case with non-abelian groups is at order
14.

\vspace{0.2cm}

\noindent $o(G_{\Phi}) = 14$

\vspace{0.2cm}

\noindent $D_7$:  $D_7$ has irreps $1, 1', 2_1, 2_2, 2_3$. There are
potentially three models, but from the symmetry of the
multiplication table they are all equivalent. So consider the $2_1$
case. Since $2_1 \times 2_1 = (1 + 2_2)_S + (1')_A$, there are two
ways to symmetrically contract $\eta^2$ and two terms of the form
$\Phi^{\dag} \cdot \Phi \eta^2$ in $V_{HFB}$. Furthermore, since
$(2_1 \times 2_1) \times 2_1 = 2_1 + 2_2 + 2_1 + 2_3$, there are no
cubic terms.

The next cases are at order 16.

\vspace{0.2cm}

\noindent $o(G_{\Phi})=16$

\vspace{0.2cm}

\noindent $(Z_4 \times Z_2) \tilde{\times}  Z_2$:  (This group is
listed as 16/8 in \cite{ThomasWood}. Here $\tilde{\times}$ is a
twisted product, as opposed to the direct product $\times$.)

This group has irreps $1_1, 1_2, ..., 1_8, 2, 2'$. Again, there is a
symmetry in the multiplication making the $2$ and $2'$ models
equivalent. Consider, therefore the case where $\Phi$ and $\eta$ are
in $2$. One has that $2 \times 2 = 1_5 + 1_6 + 1_7 + 1_8 = (1_a +
1_b + 1_c)_S + (1_d)_A$, where $a, b, c, d$ are $5, 6, 7, 8$ in some
order that will not be relevant. There are thus three terms of the
form $\Phi^{\dag} \cdot \Phi \eta^2$ in $V_{HFB}$. Furthermore,
since $(2 \times 2) \times 2 = 2 + 2 + 2' + 2'$, there can be no
cubic terms in $V_{HFB}$.

\vspace{0.2cm}

\noindent $Z_4  \tilde{\times}  Z_4$:  (16/10 in the notation of
\cite{ThomasWood}.) The irreps are $1_1, 1_2, ..., 1_8, 2, 2'$.
Again, there is a symmetry in the multiplication table and only one
independent model. Consider then the $2$, for which $2\times 2 = 1_1
+ 1_3 + 1_5 + 1_7 = (1_1 + 1_a + 1_b)_S + (1_c)_A$, where $a, b, c$
are $3, 5, 7$ in some order. We see that there will be three terms
of the form $\Phi^{\dag} \cdot \Phi \eta^2$ in $V_{HFB}$. Also since
$(2 \times 2) \times 2 = 2 + 2 + 2 + 2$, there will be no cubic
terms.

\vspace{0.2cm}

\noindent $Z_8 \tilde{\times}  Z_2$:  (16/11 in the notation of
\cite{ThomasWood}.) This case is similar the the previous example.
The irreps are again $1_1, 1_2, ..., 1_8, 2, 2'$. A symmetry in the
multiplication table leaves only one independent model. Consider
then the $2$, for which $2 \times 2 = 1_2 + 1_4 + 1_6 + 1_8 =(1_a +
1_b + 1_c)_S + (1_d)_A$, where $a, b, c, d$ are $2, 4, 6, 8$ in some
order that will not be relevant. We see that there will again be
three terms of the form $\Phi^{\dag} \cdot \Phi \eta^2$ in
$V_{HFB}$. And since $(2 \times 2) \times 2 = 2' + 2' + 2' + 2'$,
there will be no cubic terms.

\vspace{0.2cm}

\noindent $(Z_8  \tilde{\times}  Z_2)''$:  (16/13 in the notation of
\cite{ThomasWood}.) This group has irreps $1_1, 1_2, 1_3, 1_4, 2_1,
2_2, 2_3$. There is a symmetry between $2_1$ and $2_3$ so they give
the same model, but the $2_2$ is different.

(i) For the $2_1$ case one has $2_1 \times 2_1 = 1_2 + 1_3 + 2_2 =
(1_a + 2_2)_S + (1_b)_A$, and hence two terms of the form
$\Phi^{\dag} \cdot \Phi \eta^2$ in $V_{HFB}$. Here $a, b$ are $2,
3$. Since $(2_1 \times 2_1) \times 2_1 = 2_3 + 2_3 + 2_1 + 2_3$
there are no cubic terms in $V_{HFB}$.

(ii) For the $2_2$ case one has $2_2 \times 2_2 = 1_1 + 1_2 + 1_3 +
1_4 = (1_1 + 1_a + 1_b)_S + (1_c)_A$ where $a, b, c$ are $1, 2, 3$
in some order. We therefore have three terms of the form
$\Phi^{\dag} \cdot \Phi \eta^2$ in $V_{HFB}$. Since $(2_2 \times
2_2) \times 2_2 = 2_2 + 2_2 + 2_2 + 2_2$, there are no cubic terms
in $V_{HFB}$.

\vspace{0.2cm}

\noindent $D_8$: The irreps are $1_1, 1_2, 1_3, 1_4, 2_1, 2_2, 2_3$.
There is a symmetry between $2_1$ and $2_3$ so they give the same
model, but the $2_2$ is different.

(i) For the $2_1$ case, one has $2_1 \times 2_1 = (1_1 + 2_2)_S +
(1_3)_A$ and hence two terms of the form $\Phi^{\dag} \cdot \Phi
\eta^2$ in $V_{HFB}$. Since $(2_1\times 2_1) \times 2_1 = 2_1 + 2_1
+ 2_3 + 2_1$, there are no cubic terms in $V_{HFB}$.

(ii) For the $2_2$ case, one has $2_2 \times 2_2 = 1_1 + 1_2 + 1_3 +
1_4 = (1_1 + 1_a + 1_b)_S + (1_c)_A$, where $a, b, c$ are $1, 2, 3$
in some order. There are therefore three terms of the form
$\Phi^{\dag} \cdot \Phi \eta^2$ in $V_{HFB}$. Since $(2_2 \times
2_2) \times 2_2 = 2_2 + 2_2 + 2_2 + 2_2$, there are no cubic terms
in $V_{HFB}$.

We see that these two models are the same as the
two $(Z_8  \tilde{\times}  Z_2)''$ models.

\vspace{0.2cm}

\noindent $Q_8$: $Q_8$ has the same irreps and product table as
$D_8$, therefore it generates the same two models as  $D_8$ and
$(Z_8 \tilde{\times}  Z_2)''$.

\end{document}